\documentclass[12pt]{article}
\newtheorem{theorem}{Theorem}[section]
\newtheorem{proposition}[theorem]{Proposition}
\newcommand{\Z}{{\bf Z}}

\begin{document}

\title{Chaotic Properties of Subshifts \\
        Generated by a  Non-Periodic Recurrent Orbit  }

\author{}

\date{July 1999 }

\maketitle

\begin{center}

{\large   Xin-Chu Fu$^{1}$,   Yibin Fu$^{2}$,
     Jinqiao Duan$^{3}$, and  Robert S. MacKay$^{4}$   }  \\

\vspace{0.2cm}

1. {\em  Department of Mathematics and Statistics\\
        University of Surrey, Guildford, GU2 5XH, U. K. } \\
2. {\em  Department of Mathematics, Keele University \\
          Keele, Staffordshire, ST5 5BG, U. K. }\\
3. {\em  Department of Mathematical Sciences, Clemson University\\
             Clemson, South Carolina 29634, U. S. A. }\\
4. {\em   Nonlinear Centre,
     Department of Applied Mathematics and Theoretical Physics\\
     Cambridge University, Silver Street, Cambridge CB3 9EW, U. K.  }

\end{center}

\begin{abstract}

\noindent
The chaotic properties of some subshift maps are investigated.
These subshifts are the orbit closures of certain non-periodic
recurrent points of a shift map.
We first provide a review of basic concepts
for dynamics of continuous maps in metric spaces. These concepts include
nonwandering point, recurrent point, eventually periodic point,
scrambled set, sensitive dependence on initial
conditions, Robinson chaos, and topological entropy.
Next we review the notion of shift maps and subshifts.
Then we show that the one-sided
subshifts generated by a non-periodic recurrent point
are chaotic in the sense
of Robinson.  Moreover, we show that such a
subshift has an infinite scrambled set
if it has a periodic point.  Finally, we give some examples and discuss the
topological entropy of these subshifts, and present two
open problems on the dynamics of subshifts.

\vspace{0.1cm}

\end{abstract}

{\em Short Running Title:} Chaotic Properties of Subshifts

\newpage

\section{Introduction}

Periodic and chaotic behaviour are the two poles of
dynamical behaviour
of a nonlinear system. These two poles, however,
are closely related. An early
paper  discussing this topic is  Li and Yorke's
work \lq\lq Period three implies
chaos" \cite{LY}. They proved that
for a one-dimensional continuous map on an interval, if it has a periodic
point with period 3, then it is chaotic in a strong sense that they define.
Related and similar work
was also published even earlier by Sharkovskii \cite{Sharkovskii}.
There are other similar results
for higher dimensional continuous maps; see, e.g.~\cite{Marotto, LM, DM, Boy}.

Discrete dynamical systems have been used as simplified prototypical
models for some engineering and biological
processes, e.g.~\cite{May, Duan-Fu, Collet, Hao2, Strogatz}.
Through the Poincar\'e maps, discrete dynamical systems have also been used
to show ``chaos" in continuous dynamical systems, e.g.~\cite{GH, Wiggins}.

In this article, we first review basic concepts
for dynamics of continuous maps on metric spaces (Section 2).
These concepts include
nonwandering point, recurrent point, eventually periodic point,
scrambled set, sensitive dependence on initial
conditions, Robinson chaos, and topological entropy.
Then we recall the definition of shift maps and
their closed invariant subsets, called subshifts (Section 3).
Our primary interest is in one-sided subshift maps generated by
a non-periodic recurrent point.
We demonstrate that these subshifts
are chaotic in the sense
of Robinson (Section 4). Moreover we show that
such a subshift has an infinite scrambled set
if it has a periodic point (Section 5).
Finally, we give some examples and discuss their
topological entropy, and present two
open problems on the dynamics of subshifts (Section 6).

\section{Principal concepts}

In this section we review the principal concepts in the theory of
the dynamics of continuous maps on metric spaces,
and introduce related notations.

Let $F: M\rightarrow M$  be a continuous map, defined on
a metric space $ M $ with metric $\rho$.
A point $x \in M$ is called {\em nonwandering} if for any
neighbourhood $U$ of $x$, there
exists $n > 0$ such that $F^n (U) \cap U \ne \emptyset $.
A point $x \in M$ is  called a {\em periodic point} of $F$, if
$F^n(x)=x$ for some $n > 0$. The minimal such $n$ is
called the {\em period} of this periodic point.
A point $y$ is called
an {\em eventually periodic point} of $F$ if there exists $n\ge 0$ such that
$F^n(y)$ is a periodic point of $F$.
A point $x \in M$ is called a {\em recurrent point} of $F$,
if for any neighbourhood
$U$ of $x$, there exists $n > 0$ such that $F^n (x) \in U$.
The {\em $\omega$-limit set} of a point $x$ under $F$ is
\begin{equation}
\omega(x,F) = \{y\in M: \exists n_i \rightarrow \infty, F^{n_i}(x)
\rightarrow y \} .
\end{equation}

In the following, we denote by $\Omega(M,F)$, $P(M,F)$, $EP(M,F)$ and $R(M,F)$
the sets of nonwandering points, periodic points,
eventually periodic points and recurrent points, of $F$ on $M$, respectively.

There are many definitions of chaos (e.g.~see the discussions in
\cite{KS,Robinson}).  We introduce three principal ones:
\begin{enumerate}

\item{\bf Infinite scrambled set}:
Two points $a, b $ in $M$  form a {\em chaotic pair}  \cite{Zhou}
for the map $F$,
if $a$ and $b$ are
nonwandering and non-periodic points,  and
the following conditions hold:
$$ {\rm (i)} \;\; \limsup\limits_{n\rightarrow +\infty} \; \rho
(F^n (a), F^n(b))>0, $$
$$ {\rm (ii)} \;\; \liminf\limits_{n\rightarrow +\infty}\; \rho
(F^n (a), F^n(b))=0. $$
A subset $S \subseteq M$ is called a  {\em scrambled set} \cite{Smital},
if for any $a,b\in S$,
with $a\ne b$, then $(a,b)$ is a chaotic pair.
Our first notion of chaos is possession of an infinite scrambled set.
This is closely related to that used in \cite{LY}, but they did not
require chaotic pairs to be non-wandering; instead they required an
uncountable scrambled set and also a relation to $P(M,F)$
which we will not give here.  Our
feeling is that Li and Yorke's definition of chaos was of historical
importance but is not general enough.


\item{\bf Robinson chaos}:
A map $F$ is said to have {\em sensitive dependence on initial conditions}
(SDIC) on $M$
if there exists $\delta >0$ such that, for any $x\in M$
and any neighbourhood $U$
of $x$, there exists $y \in U$ and $n>0$ such that
$\rho (F^n(x),F^n(y)) > \delta $.
A map $F$ is said to be {\em topologically transitive} on $M$ if for any
two open sets $U, V  \subset M$, there exists an integer
$n>0$ such that $F^n(U) \cap V \neq \emptyset$.
An equivalent definition is that there is a dense orbit.
This condition might not apply directly to $F$ but often applies to the
restriction of $F$ to some closed invariant subset.
A map is called {\em chaotic in the sense of Robinson} \cite{Robinson}
if it has sensitive dependence on initial conditions
and is topologically transitive.
This definition is based on one of Devaney \cite{Devaney}, who also
required the periodic points to be dense, but as argued by Robinson,
this does not seem to be intrinsic to the phenomenon of chaos, so we
leave it out.  A curious twist, incidentally, is that Banks et al
\cite{Banks} realised that topological transitivity plus density of
periodic points imply SDIC, so
that SDIC can be left out of Devaney's definition with making any change
to his concept.  Finally, note that
Wiggins \cite{Wiggins} defines chaos to mean SDIC plus
$M$ has more than one orbit.

\item{\bf Positive topological entropy}:

Topological entropy \cite{Adler, Alseda, Gutzwiller}
was first defined by C. E. Shannon \cite{Shannon} in 1948     and
called by him {\em noiseless channel capacity}.
Engineers prefer this term while
mathematicians like the term {\em topological entropy}.
It is an important quantity measuring the rate of growth of the
complexity of the orbit structure of a
dynamical system with respect to time.
In the case where $M$ is compact, which suffices for this paper, it is
most simply defined as follows.  Given $\epsilon > 0$, say two orbit
segments $(x_0,\ldots,x_n)$ and $(y_0,\ldots,y_n)$ of length $n > 0$
are {\em $\epsilon$-distinguishable} if there exists $j \in
\{0,\ldots,n\}$ such that $d(x_j,y_j) > \epsilon$.  Let $N(n,\epsilon)$
be the maximum size of a set of $\epsilon$-distinguishable orbit
segments of length $n$ (which is finite by compactness).  Then the
{\em topological entropy} of $F$ is
\begin{equation}
ent(F) = \lim_{\epsilon \rightarrow 0} \limsup_{n\rightarrow \infty}
\frac{1}{n} \log N(n,\epsilon) .
\end{equation}
We say $F$ has {\em topological chaos} if its topological entropy is
positive.


\end{enumerate}


\section{Shift Maps and Subshifts}

We now recall the definitions of shift maps and subshifts,
and define the class of subshifts to be studied
in this paper.

Let $ X $ be a metric space with metric $d$.
The basic example to bear in mind is $X = \{0,\ldots,N-1\}$
(often denoted just by $N$) for some
integer $N>1$ with the discrete metric:
$$ d(m,n)=\left\{\begin{array}{ll} 0, \;\;\;\; m=n\\ 1, \;\;\;\; m\ne n .
 \end{array} \right. $$
Denote by $\Sigma^X$ the space
consisting of one-sided sequences in the
metric space $X$ (it can alternatively be written as $X^{\Z_+}$). So
$x \in \Sigma^X$ may be denoted by $x=(x_0, x_1, \dots, x_i, \dots)$,
$x_i \in X$, $i \ge 0$. Let $\Sigma^X$ be endowed with the product
topology. Then $\Sigma^X$ is metrizable, and the metric on $\Sigma^X$ can
be chosen to be
$$ \rho(x, y)=\sum^{+\infty}_{i=0} \frac{1}{2^i} \frac{d(x_i, y_i)}{1+
d(x_i, y_i)}, \;\;\;\; x=(x_0, x_1, \dots), \;\; y=(y_0,y_1, \dots) \in
\Sigma^X.
$$

The {\em shift map} $\sigma$: $\Sigma^X \rightarrow \Sigma^X$ is defined by
$(\sigma(x))_i=x_{i+1}, \; i=0,1, \cdots.$
Since
$$ \rho (\sigma(x), \sigma(y)) \le 2 \rho(x,y), $$
$\sigma$ is continuous.
If $X$ has more than one element, the shift map is chaotic in all three
senses that we have defined.
A two-sided shift map can be defined similarly, acting by the same
formula but on doubly infinite sequences of elements of $X$.  We shall
consider only one-sided shifts in this paper.

As an aside, when $X$ is a vector space,
$\Sigma^X$ becomes a vector space in the natural way.
If $X$ is a nontrivial vector space, $\Sigma^X$ is infinite dimensional.
Fu and Duan \cite{Fu-Duan} have shown that the shift map
$\sigma$ induces an infinite dimensional {\em linear chaotic}
discrete-time system. Furthermore, this shift map has been used to demonstrate
the chaotic behaviour of a quantum harmonic oscillator
by Duan et al \cite{Duan-Fu}.

A closed $\sigma$-invariant subset of $\Sigma_X$ is called a {\em
subshift}.  Restricting attention now to the case of a set $X$ with
the discrete topology, a subshift is said to be {\em of finite type} if it can
be completely
specified by giving a finite list of excluded words, i.e. finite strings of
elements of $X$ that do not occur in any of the sequences of the
subshift.  The topological entropy of a shift or subshift with $X$ discrete
is equal to the exponential growth rate of the number of
different blocks of length $n$ occurring in the symbol sequences.
For a shift this is just $\log N$, where $N$ is the size of $X$.  For
subshifts of finite type, it reduces to computing the largest
eigenvalue of an associated transition matrix, and is positive if
the subshift does not consist only of periodic and eventually periodic
points.  In this case, the subshift of finite type is also chaotic in
the other two senses.
We call shifts and subshifts {\em symbolic dynamics systems}.
Symbolic dynamics is a powerful tool to study
more general dynamical systems, because the latter often contain
invariant subsets on which the dynamics is equivalent to a shift
map or subshift.
Readers are referred to a recent
book \cite{Kitchens} for more background and details on symbolic dynamics.


For the rest of the paper, we will concentrate on subshifts that arise
by taking the closure of a single orbit of a shift.  These are not
necessarily subshifts of finite type, and this is why their study is
interesting and non-trivial.
The study of orbit
closures goes back at least to Morse, who studied the closure of the
famous Morse sequence under the shift,
which yields what we now call an ``adding machine''.
We refer to \cite{Gottschalk} for an early source discussing orbit closures.
Denote by $\Sigma(a)$ the closure of the orbit starting at point
$a \in \Sigma^X$, i.e.,
$$ \Sigma(a)=cl (\left\{\sigma^n(a):\; n\ge 0 \right\}). $$

It is easy to show that $\sigma(\Sigma(a)) \subseteq \Sigma(a)$,
and it is closed by construction. So
$(\Sigma(a), \sigma)$ is a subshift.
We call $(\Sigma(a), \sigma)$ the subshift generated
by the orbit starting at point $a$, or, in short, by the point $a$.

In the following, we discuss dynamical behaviour of the subshift
$(\Sigma(a), \sigma)$.

\section{Properties of Subshift $(\Sigma(a), \sigma)$}

We now discuss the dynamical properties of the subshift $(\Sigma(a),
\sigma)$ generated by a point $a \in \Sigma^X$.
Initially, $X$ is an arbitrary metric space, but we shall rapidly
specialise to the case $X = N$.  The results of this section are
elementary but we spell them out for pedagogical purposes.

Suppose $x \in \Sigma^X$ is recurrent but non-periodic, i.e.,
$x \in R(\Sigma^X, \sigma)-P(\Sigma^X, \sigma)$. Since $x \in
R(\Sigma^X, \sigma)$, there exist $\{ n_i\}$, $n_i \rightarrow +\infty$ as $i
\rightarrow +\infty $, such that when $i$ is big enough, $\sigma^{n_i}(x)$
enters any neighbourhood of $x$. Since $x \notin P(\Sigma^X, \sigma)$,
$\sigma^{n_i}(x) \ne x$, $\forall i \ge 1$. So ${n_i}$ can be chosen such
that $\sigma^{n_i}(x)$, $i=1, 2, \dots, $ are all different, and thus
$\{\sigma^{n}(x), n\ge 0 \}$  is a countably infinite set.
This implies that $x \notin
EP(\Sigma^X, \sigma)$.
So we have the following result

\begin{proposition}\label{prop31}
Non-periodic  recurrent points of $\sigma$ are not eventually periodic points,
i.e.,
   $(R(\Sigma^X, \sigma)-P(\Sigma^X, \sigma))
  \cap  EP(\Sigma^X, \sigma)=\emptyset$.
\end{proposition}

Now suppose $a \in R(\Sigma^N, \sigma)-P(\Sigma^N, \sigma)$. Because
$\Sigma(a)$
is a closed subset of the compact space $\Sigma^N$, $\Sigma(a)$ is compact.
Also, since $\Sigma^N$ is totally disconnected, its subset
$\Sigma(a)$ is also totally disconnected. From $a \in R(\Sigma^N,\sigma)$, we
have $\sigma^n(a) \in R(\Sigma^N, \sigma)$, $\forall n\ge 0$. By
Proposition~\ref{prop31},
$\sigma^n(a) \notin P(\Sigma^N, \sigma)$. Therefore, $\sigma^n(a)
\in R(\Sigma^N, \sigma)-P(\Sigma^N, \sigma)$, $\forall n \ge 0$. So there
exist $\{n_i \}, n_i \rightarrow +\infty$ as $ i \rightarrow +\infty$, such
that $\sigma^{n_i}(\sigma^n(a))$, $i=1,2, \dots$ are all different, and
$\sigma^{n_i}(\sigma^n(a))$ enters any neighbourhood of $\sigma^n(a)$. These
imply $\sigma^n(a) \in (\{ \sigma^m(a), m\ge 0\})'$, $\forall n \ge 0$, where
for a subset $S$ of a topological space, $S'$ denotes its set of limit
points. So
$\{ \sigma^n(a), n\ge 0\} \subseteq (\Sigma(a))'$. So
$(\Sigma(a))'=\Sigma(a)$, i.e., $\Sigma(a)$ is perfect, and $\Sigma(a)$
is homeomorphic to a Cantor set \cite{Wiggins}.
We summarize this result as

\begin{proposition}\label{prop32}
 If   $a$ is a
non-periodic  recurrent point for $\sigma$, i.e.,
    $a \in R(\Sigma^N, \sigma)-P(\Sigma^N, \sigma)$,
then $\Sigma(a)$ is homeomorphic to the Cantor set.
\end{proposition}

In particular, if $a \in R(\Sigma^N, \sigma)-P(\Sigma^N, \sigma)$, then
$\Sigma(a)$
is uncountable. Moreover, for $a \in R(\Sigma^N, \sigma)$, we have
$\Sigma(a) =\omega(a, \sigma)$ (the $\omega$-limit set of the orbit from $a$).
The converse is also true. So
$a \in R(\Sigma^N, \sigma)$ if and only if $\Sigma(a) =\omega(a, \sigma)$.

We further have the following conclusion
\begin{proposition}\label{prop33}
 If   $a$ is a
non-periodic  recurrent point for $\sigma$, i.e.,
$a \in R(\Sigma^N, \sigma)-P(\Sigma^N, \sigma)$,
then $\sigma$ has sensitive dependence on initial conditions on
$\Sigma(a)$.
\end{proposition}

This can be shown as follows. Take $\delta_0=1/2$.
For $ x \in \Sigma(a)$,
let $V$ be an arbitrary neighbourhood of $x$ in
$\Sigma(a)$. If there exists $y \in V$, such that $y \ne x$, then taking
$k=\min\{n\ge 0: x_n\ne y_n\}$, we have
$$\rho(\sigma^k(x), \sigma^k(y))=d(x_k,y_k)+\cdots \ge 1 > \delta_0.$$
So we only need to show the existence of $y$.

If there exists $n$ such that
$x=\sigma^n(a)$, then from Proposition~\ref{prop32}, $x \in R(\Sigma^N,
\sigma)-P(\Sigma^N, \sigma)$. Therefore, there exist
$\{ n_i\}\rightarrow \infty$, such that
$\sigma^{n_i}(x)$, $i=1,2, \dots$, are all different, and
$\lim_{i\rightarrow +\infty} \sigma^{n_i}(x)=x$. So we can take
$y=\sigma^{n_i}(x)$ ($i$ large enough).
Alternatively, if $\sigma^n(a) \ne x$ for all $n \in \Z_+$ then
there exist $\{ m_i\}$, $m_i \rightarrow \infty$,
such that $x =\lim_{i\rightarrow +\infty}
\sigma^{m_i}(a)$, and hence we can take
$y = \sigma^{m_i}(a)$ for $i$ large enough.

This completes the proof of Proposition~\ref{prop33}.
The important ingredient was that the shift on one-sided sequences is
{\em expanding}, meaning that there exists $\delta > 0$ such that for
all pairs of distinct points $x,y$ there exists $n \ge 0$ such that
$\rho(\sigma^n(x),\sigma^n(y)) \ge \delta$.  The result does not apply to
the two-sided shift.

We now discuss the image and nonwandering set of the
subshift $\sigma$.
Let $a$ be a recurrent point, i.e.,
$a\in R(\Sigma^X, \sigma)$.  It is easy to verify that $\sigma(\Sigma(a))
=\Sigma(a)$.  Since the orbit of $a$ is dense in $\Sigma(a)$,
$\sigma$ is topologically transitive on $\Sigma(a)$. Thus every point
is nonwandering.
We thus obtain the following conclusion.
\begin{proposition}\label{prop34}
 Assume that $a$ is a recurrent point, i.e.,
$a \in R(\Sigma^X, \sigma)$. Then
$\sigma(\Sigma(a))=\Sigma(a)$, $\Omega(\sigma|_{\Sigma(a)})=\Sigma(a)$.
\end{proposition}

For $a\in R(\Sigma^N, \sigma)-P(\Sigma^N, \sigma)$, Proposition~\ref{prop33}
claims that
$\sigma$ has sensitive dependence on initial conditions on $\Sigma(a)$.
As noted in the proof of  Proposition~\ref{prop34},
$\sigma$ is topologically transitive on $\Sigma(a)$. So we have
\begin{theorem}\label{th35}
If $a$ is a
non-periodic  recurrent point for $\sigma$, i.e.,
$a \in R(\Sigma^N, \sigma)-P(\Sigma^N, \sigma)$,
then the subshift $(\Sigma(a), \sigma)$ is chaotic in the sense of
Robinson.
\end{theorem}

\section{Further Properties of Subshift $(\Sigma(a), \sigma)$
	When It Has a Periodic Point}

The above discussion shows some interesting properties of
subshifts generated by
non-periodic recurrent points. In particular, Theorem~\ref{th35} declares
that all these subshifts
are chaotic in the sense of Robinson. In this section we will show
that if $\Sigma(a)$
contains a periodic point, then there
exist infinitely many chaotic pairs in
$\Sigma(a)$; in particular, if $\Sigma(a)$ contains a fixed point, then there
exists an infinite scrambled set which is dense in $\Sigma(a)$.
\begin{theorem}\label{th41}
Assume that $a$ is a
non-periodic recurrent point and that $\Sigma(a)$ contains a
periodic point of $\sigma$, i.e.,
$a \in R(\Sigma^N, \sigma)-P(\Sigma^N,
\sigma)$ and $\Sigma(a) \cap P(\Sigma^N, \sigma) \ne \emptyset$.
Then there exists
an infinite set $S \subset \Sigma(a)$ in which any two different points are
a chaotic pair for the subshift $\sigma$,
i.e., $S$ is an infinite scrambled set for
the subshift $\sigma$.
\end{theorem}
\noindent {\bf Proof}: Take $x_0 \in \Sigma(a) \cap P(\Sigma^N, \sigma)$ and
denote by $k$ the period of $x_0$. Define $S=\{\sigma^{kn} (a), \; n\ge 0 \}$.
{}From Proposition~\ref{prop34}, $S \subseteq \Omega(\sigma|_{\Sigma(a)})$. By
Proposition~\ref{prop31},
$S \cap P(\Sigma^N, \sigma) =\emptyset$. That is, every point in $S$ is a
nonwandering  and non-periodic point.

Whenever $m_1 \ne m_2$, there exists a subsequence $\{n_i \}$ such that
$(\sigma^{km_1}(a))_{n_i} \ne (\sigma^{km_2}(a))_{n_i}$, where
subscript $n_i$ denotes the $n_i$-entry.
For suppose that this claim is not true. Then there would exist
$N>0$,
such that $\sigma^{N} (\sigma^{km_1}(a))=\sigma^{N} (\sigma^{km_2}(a))$.
Without loss of generality, we assume that $m_1 <m_2$. Thus
$$
\sigma^{N+km_1}(a)=\sigma^{N+km_2}(a)=\sigma^{k(m_2-m_1)}(\sigma^{N+km_1}(a)).
$$ So $a \in EP(\Sigma^N, a)$. That is a contradiction.

Therefore, $S$ is an infinite set, and there exist $\{ n_i\}, \;
n_i \rightarrow +\infty$ as $i \rightarrow +\infty$, such that
$$ \rho\left(\sigma^{n_i}(\sigma^{km_1}(a)), \;
\sigma^{n_i}(\sigma^{km_2}(a)) \right) \ge 1, $$
i.e.,
$$
\limsup\limits_{n\rightarrow +\infty} \; \rho (\sigma^n(x),
\sigma^n(y)) \ge 1 >0, \;\;\;\; \forall x, y \in S, \;\; x \ne y.$$
This proves condition (i) in the definition of a chaotic pair
in Section 1.

Since $x_0 \in \Sigma(a)=\omega(a, \sigma)$,
there exists a subsequence $\{ n_i\}$ such that
$$ \lim_{i\rightarrow +\infty} \rho (\sigma^{n_i}(a), x_0)=0, $$
i.e., $\forall M>0, \; \exists N>0$, such that whenever $i \ge N$, we have
$$
\rho (\sigma^{n_i}(a), x_0)< \frac1{2^{Mk}}.
$$
Hence the first $Mk+1$ entries of $\sigma^{n_i}(a)$ and $x_0$ agree.
Therefore,
$\forall 0 \le m <M$, whenever $i\ge N$, we have
$$
\rho (\sigma^{n_i}(\sigma^{km}(a)), x_0)\le  \frac1{2^{(M-m)k} }.
$$
This implies that for $\forall m \ge 0$, we have
$$ \lim_{i\rightarrow +\infty} \rho (\sigma^{n_i}(\sigma^{km}(a)), x_0)=0.$$
Moreover, because
$$ \rho \left(\sigma^{n_i}(\sigma^{km_1}(a)),
\sigma^{n_i}(\sigma^{km_2}(a)) \right) \le \rho
\left(\sigma^{n_i}(\sigma^{km_1}(a)), x_0 \right)+
\rho \left(\sigma^{n_i}(\sigma^{km_2}(a)), x_0 \right), $$
we have
$$ \lim_{i\rightarrow +\infty}  \rho \left(\sigma^{n_i}(\sigma^{km_1}(a)),
\sigma^{n_i}(\sigma^{km_2}(a)) \right)=0, \;\; \forall m_1, m_2 \ge 0,$$
i.e.,
$$
\liminf\limits_{n\rightarrow +\infty} \; \rho (\sigma^n(x), \; \rho
(\sigma^n(y))=
0, \;\;\; \forall x, y \in S, \;\; x \ne y.$$
This proves condition (ii) in the definition of a chaotic pair
in Section 1.
So any pair of distinct points $x, y$ in $S$ is a chaotic pair.
This completes the proof of Theorem~\ref{th41}.

We remark that if $\sigma$ has a fixed point
(a periodic point with period $k=1$), the chaotic pair set $S$ is
even more special. Recall the construction of $S$ in
the proof of Theorem~\ref{th41},
$S=\{\sigma^{kn}(a), \; n\ge 0 \}$, where $k$ is the period of some periodic
point in $\Sigma(a) \cap P(\Sigma^N, \sigma)$. If
$\Sigma(a) \cap P(\Sigma^N, \sigma)$ contains a fixed point of $\sigma$,
then we
can take $k=1$, so $S=\{\sigma^{n}(a), \; n\ge 0 \}$. Therefore $S$ is dense in
$\Sigma(a)$ and $\sigma(S) \subseteq S$. So we have
\begin{theorem}\label{th42}
 Assume that $a$ is a
non-periodic recurrent point and that $\Sigma(a)$ contains a
fixed point
(a periodic point with period $1$) of $\sigma$.
  Then the subshift $\sigma$ has a dense
invariant scrambled     set $S \subseteq \Sigma(a)$.
\end{theorem}

\section{Examples, Discussions and Open Problems}

The subshifts of the form $(\Sigma(a), \sigma)$ are common and fundamental,
because the full shift $(\Sigma^N, \sigma)$ and any subshift of finite type
$(\Sigma_A, \sigma_A)$ with an irreducible transition matrix $A$ can be
generated by a point in $R(\Sigma^N, \sigma)-P(\Sigma^N,\sigma)$.  In
particular, any topologically transitive subshift can be written in the form of
$(\Sigma(a), \sigma)$, and by definition a chaotic subshift should be
topologically transitive.  Also it is obvious that any subshift contains
subshifts of the form $(\Sigma(a), \sigma)$.

Moreover, many subshifts of infinite type can be generated by points
in the subset $R(\Sigma^N, \sigma)-P(\Sigma^N,\sigma)$.
The following are three examples.

\vspace{0.2cm}

\noindent {\bf Example 6.1.} (\cite{FW}) Given
sequences $s_i$ of $0$ and $1$ as
follows: $s_0=00110$, $s_1=t_0s_0$, $s_2=t_1s_0s_0s_1$, \dots,
$s_n=t\overbrace{s_0 \cdots s_0}^n s_1s_2 \cdots s_{n-1} (n\ge 2)$,
\dots, where $t_0=00000$, $t_1=11111$, $t=t_0$ when $n$ is odd, and
$t=t_1$ when $n$ is even, take $a=(s_0s_1s_2\cdots s_n\cdots)\in \Sigma^2$,
then $a\in $ $R(\Sigma^2, \sigma)-P(\Sigma^2,\sigma)$.
$(\Sigma(a), \sigma)$ is a subshift of infinite type. $\Sigma(a)$ is
homeomorphic to the Cantor set; $\Omega(\sigma|_{\Sigma(a)})=\Sigma(a)$;
$(\Sigma(a), \sigma)$ is chaotic in the sense of Robinson. Because
the
$5$-period point $(s_0s_0 \cdots s_0\cdots)\in \Sigma(a)$,
$S_1=\{\sigma^{5n}(a), n\ge 0 \}$ is an infinite scrambled set (from
Theorem~\ref{th41}).

\vspace{0.2cm}

\noindent {\bf Example 6.2.} (\cite{FW}) Take
sequences $s_0=00110$, $t_1=11111$. Form
a symbol sequence $b\in \Sigma^2$ by concatenating all
words  of  finite length formed from
 $s_0$ and $t_1$. Then $b\in R(\Sigma^2, \sigma)-P(\Sigma^2,\sigma)$.
$(\Sigma(b), \sigma)$ is a subshift of infinite type. $\Sigma(b)$ is
homeomorphic to the Cantor set; $\Omega(\sigma|_{\Sigma(b)})=\Sigma(b)$;
$(\Sigma(b), \sigma)$ is chaotic in the sense of Robinson.
Because the fixed
point $(11\cdots1\cdots)\in \Sigma(b)$, $S_2=\{\sigma^n(b), n\ge 0 \}$ is a
dense
invariant scrambled set (according to Theorems~\ref{th41} and~\ref{th42}).
\vspace{0.2cm}

\noindent {\bf Example 6.3.} (\cite{Zhou}) Let
$P=i_0i_1\cdots i_{n-1}$ and
$Q=j_0j_1\cdots j_{m-1}$ be words of $0$ and $1$ with lengths
$|P|=n$ and $|Q|=m$. Take $P_0=10$, $Q_0=00$, $P_1=P_0Q_0$, $Q_1=0\cdots0$,
$|Q_1|=|P_0Q_0P_1|\times 2=|P_1P_1|\times 2$. For $n>1$, let
$P_n=P_0Q_0P_1Q_1\cdots P_{n-1}Q_{n-1}=P_{n-1}P_{n-1}Q_{n-1}$,
$Q_{n-1}=0\cdots0$, $|Q_{n-1}|=|P_0Q_0\cdots P_{n-1}|\times n=|P_{n-1}P_{n-1}
| \times n$. Take $c=(P_0Q_0P_1Q_1\cdots P_nQ_n\cdots)\in \Sigma^2$. Then
$c\in  R(\Sigma^2, \sigma)-P(\Sigma^2, \sigma)$, and $(\Sigma(c), \sigma)$ is
a subshift of infinite type. $\Sigma(c)$ is homeomorphic to the Cantor set;
$\Omega(\sigma|_{\Sigma(c)})=\Sigma(c)$; $(\Sigma(c), \sigma)$ is chaotic in
the sense of Robinson. Because $\Sigma(c)\cap P(\Sigma^2, \sigma)$$=
\{e \}$, where $e=(0\cdots 0 \cdots) \in F(\Sigma^2, \sigma)$,
$S_3=\{\sigma^n(c), n\ge 0 \}$ is a dense invariant scrambled set.

Now we discuss the topological entropy
of the above subshifts.
For the subshifts constructed in Examples 6.1-3, we can calculate their
topological entropies  \cite{FW}:
$$ {\rm ent}(\sigma|_{\Sigma(a)})={\rm ent}(\sigma|_{\Sigma(b)})=
\frac{1}{5}\log 2>0,
\;\;\;\;
{\rm ent} (\sigma|_{\Sigma(c)})=0. $$
It is somewhat surprising that $(\Sigma(c), \sigma)$ has zero entropy, despite
having both Robinson chaos and an infinite scrambled set.
One may ask if there exists a subshift of
infinite type with positive topological entropy but no infinite scrambled set
nor Robinson chaos.
Of course, there are many
examples of the latter, because ${\rm ent} (\sigma|_{\Sigma})>0 $
does not imply that
$(\Sigma,\sigma)$ is topologically transitive, but we could ask instead for
a closed invariant subset with Robinson chaos.
For one- and two-dimensional smooth maps, \cite{Katok} shows that
positive entropy implies existence of a
subshift of finite type, but we do not know the answer in the general context.

Finally, we present two open problems which are closely related to the
topics of this paper:

\vspace{0.2cm}

\noindent {\bf Problem 1.} Suppose $a \in R(\Sigma^N, \sigma)-P(\Sigma^N,
\sigma)$, $\Sigma(a) \cap P(\Sigma^N, \sigma)\ne \emptyset$. Take $p\in
\Sigma(a) \cap P(\Sigma^N, \sigma)$, and denote its period by $k$. Does the
set $cl \{ \sigma^{kn}(a), n\ge 0\}$ contain an uncountable scrambled set?

\noindent {\bf Problem 2.} Can Theorem 5.1 be strengthened to give an
uncountable scrambled set?

\section{Acknowledgements}

X. Fu would like to thank Keele University and Heriot-Watt University
for financial support and
hospitality while earlier versions of this paper were written.
J. Duan would like to thank Hiroshi Kokubu, Kyoto University, Japan,
for pointing out a paper by J. Smital \cite{Smital2}.
We would like to thank P.L.Boyland and the
referees for very useful comments.

This work was partially supported by China NSF Grant No.19572075 and
UK EPSRC Grant GR/M36335  for X. Fu, and by the
National Science Foundation of USA Grant DMS-9704345 for J. Duan.


\end{document}